\documentclass{aastex}
\usepackage{emulateapj5}

\def\mic{$\mu$m\/}

\slugcomment{Accepted for publication in the Astrophysical Journal}

\shorttitle{A Submillimeter Survey of Gravitationally Lensed Quasars}
\shortauthors{Barvainis \& Ivison}

\begin{document}

\title{A Submillimeter Survey of Gravitationally Lensed Quasars}

\author{Richard Barvainis}
\affil{National Science Foundation, Arlington, VA 22230}
\affil{4201 Wilson Boulevard, Arlington, VA 22230 USA}
\email{rbarvai@nsf.gov}
\and 
\author{Rob Ivison}
\affil{Royal Observatory, Edinburgh}
\affil{Blackford Hill, Edinburgh, EH9 3HJ UK}
\email{rji@roe.ac.uk}

\begin{abstract}
Submillimeter (and in some cases millimeter) wavelength continuum
measurements are presented for a sample of 40 active galactic nuclei
(probably all quasars) lensed by foreground galaxies.  The object of
this study is to use the lensing boost, anywhere from $\sim 3- 20$ times,
to detect dust emission from more typical AGNs than the extremely luminous
ones currently accessible without lensing.  The sources are a mix of
radio loud and radio quiet quasars, and, after correction for synchrotron
radation (in the few cases where necessary), 23 of the 40 (58\%) are
detected in dust emission at 850\mic ; 11 are also detected at 450\mic .
Dust luminosities and masses are derived after correction for lensing
magnification, and luminosities are plotted against redshift from $z =
1$ to $z = 4.4$, the redshift range of the sample.  The main conclusions
are (1) Monochromatic submillimeter luminosities  of quasars are, on
average, only a few times greater than those of local IRAS galaxies;
(2) Radio quiet and radio loud quasars do not differ significantly
in their dust lumimosity; (3) Mean dust luminosities of quasars and radio
galaxies over the same redshift range are comparable; (4) Quasars and
radio galaxies alike show evidence for more luminous and massive dust
sources toward higher redshift, consistent with an early epoch of 
formation and possibly indicating that the percentage
of obscured AGNs increases with redshift.

\end{abstract}


\section{Introduction}

 Strong dust emission at infrared and submillimeter wavelengths is
now recognized as a fundamental tracer of activity in galaxies, be it
star formation or processes related to nuclear black holes.  Ordinary,
quiescent galaxies like the Milky Way exhibit dust luminosities of order
$10^{10} L_{\sun}$, but active systems, such as ultraluminous infrared
galaxies (ULIRGs), quasars, and radio galaxies can be as much as $10^3$
times more powerful (Sanders \& Mirabel 1996).  In these objects the
intense dust emission appears to be triggered by galaxy interactions and
mergers, which are thought to be responsible at least in part for the
formation and growth of massive elliptical galaxies and the supermassive
black holes found in their nuclei (e.g., Kormendy \& Sanders 1992;
van Dokkum et al.\ 1999).

  In the mid-1980s the IRAS mission discovered many infrared-luminous
galaxies and quasars at low redshifts in the mid- to far-infrared
(12--100\mic ), establishing dust emission as an important if not
dominant component of the bolometric luminosity in certain classes
of objects.  The dust luminosity is a result of reprocessing of
shorter wavelength optical/UV radiation originating either from
hot young stars, in the case of starbursts, or from the release of
gravitational energy as material accretes onto a nuclear black hole,
in the case of quasars.  In some galaxies both starburst and quasar
activity are likely to contribute, and there remains some dispute
over which process dominates in various types of objects.

  Although the findings of IRAS were revolutionary for the local universe,
its sensitivity was such that only a few high-redshift systems were
detected: the quasar PG 1634+607 ($z = 1.3$), the ULIRG/buried quasar
IRAS F10214+4724 ($z = 2.3$), the Cloverleaf quasar (H 1413+117; $z =
2.6$), and the $z = 3.9$ broad absorption line quasar APM 08279+5255
(Rowan-Robinson et al.\ 1991; Barvainis et al.\ 1995; Lewis et al.\ 1998).
The latter three are strongly magnified by gravitational lensing, boosting
their fluxes above the IRAS detection threshholds at 60 and 100\mic
. Further progress was made at submillimeter wavelengths at the James
Clerk Maxwell Telescope (JCMT), first using the single-channel photometer
UKT14, then later with the submillimeter camera SCUBA.  F10214+4724 and
the Cloverleaf were detected at 800 and 450\mic\ (Rowan-Robinson et al.\
1993; Barvainis et al.\ 1992), to be followed by several other quasars and
radio galaxies (Hughes et al.\ 1997).  To date the census of published
submillimeter detections of high redshift objects numbers $\sim 150$
(mostly unidentified), some of which are either strongly lensed by
individual galaxies or weakly lensed by galaxy clusters (Smail, Ivison, \& 
Blain 1997).

\begin{figure*}
\figurenum{1}
\epsscale{1.2}
\plotone{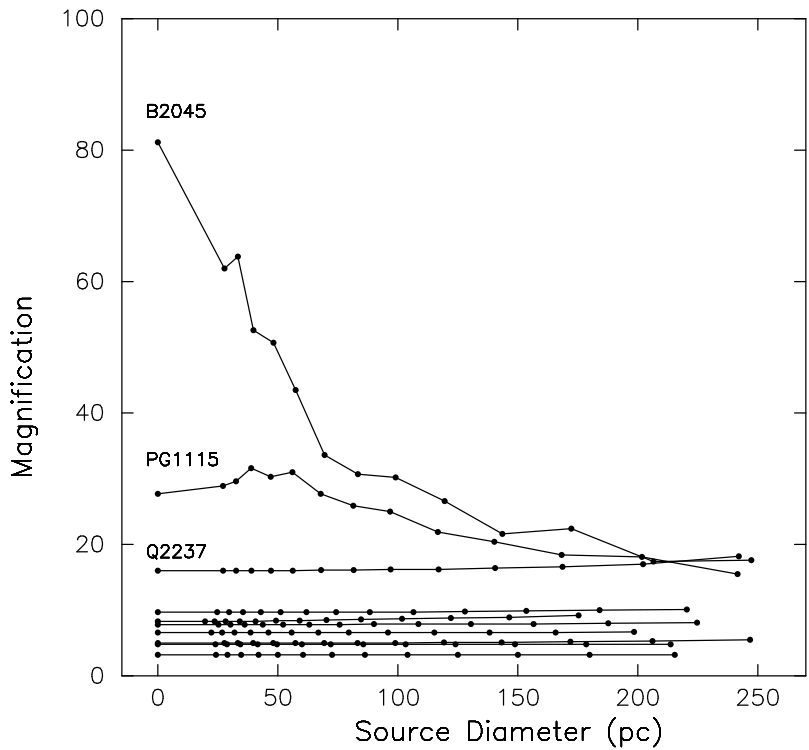}
\caption{Modeled magnification versus source diameter for some of the lensed quasars 
in the sample.  The model for the intervening lens, based on optical data, 
is applied to successively larger sources at the quasar to determine the 
net magnification.  Magnifications that are high for the optical 
emission (essentially a point source) tend to drop to about 20 
as the source size increases to $\sim 200$ pc (a likely minimum dust 
source size).  At low magnifications the magnification tends to be constant 
with source size (unlabelled examples plotted include SBS 0909+532, 
BRI 0951+2635, LBQS 1009$-$0252, HST 12531$-$2914, HST 14113+5211, H 1413+117, 
and HST 14176+5226).  Model results generated by B.  McLeod (private 
communication).
} \end{figure*}

  The high sensitivity of SCUBA, combined with the flux boost
provided by strong lensing, presents an opportunity to delve 
deeper into the submillimeter luminosity function than would
otherwise be possible.  The majority of unlensed quasars are
undetectable with current instruments, so up to now {\it typical}
dust luminosties and masses of active galaxies at high redshift
have remained unknown. Only the very highest luminosities can be
detected without lensing.  However, there are now about 50 known
strongly lensed objects, most of which are in fact quasars (see the CASTLeS
compilation, http://cfa-www.harvard.edu/castles).  This paper
reports the results of a submillimeter survey of a large fraction of the
known lensed quasars, presenting data on a sample of 40 objects
in which dust emission has been clearly detected at 850 and/or
450\mic , or meaningfully constrained by upper limits.

  These lensed quasars, while not a complete sample by
any definitive criteria, are nevertheless a random, passively
selected sample of quasars at high redshift -- they are quasars
that happen to lie behind massive galaxies.  With a typical lensing
boost of an order of magnitude, they provide a unique window on
the dust content of active galaxies in the early universe.

\section{Observations and Results}

\subsection{Data}

Of the submillimeter results on 40 lensed quasars listed in Table 1, 39
are from our own observations and one (F10214+4724) is from Rowan-Robinson
et al.\ (1993).  APM 0879+5255 was originally detected by Lewis et al.\
(1998), but our own higher SNR observations are used here.  Table 1
also reports 3mm and 1.3mm fluxes for a number of sources.  In radio
loud objects these longer wavelength data are useful for correcting
synchrotron contamination of the submillimeter fluxes to derive an
estimate of the true dust emission level.

The submillimeter observations were made over a series of observing
runs at the JCMT during 1999-2001.
The submillimeter camera SCUBA (Holland
et al.\ 1999) was used in photometry mode, observing simultaneously
at 450\mic\ and 850\mic , with the target centered on the central
bolometer of SCUBA's back-to-back arrays.  The JCMT's secondary
mirror was chopped azimuthally by 60\arcsec\ at 7\,Hz while jiggling
through a 9-point filled-square pattern, the points separated
by 2\arcsec.  The telescope was nodded in azimuth by 60\arcsec\ following
the completion of each jiggle pattern, in an object-sky-sky-object
pattern designed to keep the source constantly in one of the chop
beams.  Atmospheric opacity, pointing accuracy, and focus position
were checked every $30-60$\,min and calibration scans of Uranus
or Mars were obtained each night.   Data were analysed using {\sc
surf} (Jenness \& Lightfoot 1998) following the standard recipes
(Stevens, Ivison, \& Jenness 1997).

RMS errors for the SCUBA
data range from 0.9 to 3.6 mJy ($1\sigma$) at 850\mic , depending on
time spent on source and weather conditions.  At 450\mic\ atmospheric
transmission is much poorer than at 850\mic , and meaningful fluxes
or upper limits at this wavelength were obtained only during the
best conditions.

All 3mm continuum fluxes are from the IRAM Plateau
de Bure Interferometer (PdBI), and all but one were obtained by us
during a search for CO emission from lensed quasars (Barvainis, Alloin,
\& Bremer 2002;  the 3mm flux
for 0957+561 is from Planesas et al.\ 1999).  The 1.3mm data were
obtained either with the IRAM 30m telescope bolometer, with the PdBI,
or with the 1.3mm channel on SCUBA (see notes to Table 1).

Upper limits given in Table 1 are 2.5 times the $1\sigma$
statistical uncertainties.  Quoted errors do not include
absolute calibration uncertainties, estimated to be in the range
$5-10\%$.

\subsection{Lensing Magnifications}

In order to calculate source luminosities, lensing must be taken
into account.  Estimates of the lensing magnifications were obtained
for 30 of the 40 sources, either from the literature or via private
communication. These magnifications are generally derived from either
optical or radio images, and refer to a very compact source of emission.
When such a source is very close to a lensing caustic, the magnification
can be quite high -- over 100 in extreme cases.  In the optical, emission
is from the primary quasar optical/UV source which has a size scale of
order $10^{-3}$ pc.  Boosting derived from radio images often refers to
a flat-spectrum core, also on scales of well under a parsec.  

The basic problem in estimating the magnifications of the dust sources is
that the size scales for submillimeter dust emission are tens to hundreds
of parsecs rather than
subparsec, and sources on these scales will tend to have lower overall
magnifications, particularly when the optically-derived magnifications
are high (i.e., when the optical source is very close to a caustic).
A large source will have only a small fraction of its area near enough to
a caustic to be highly magnified, with other portions farther away and
therefore magnified less.  This ``chromatic" lensing effect, sometimes
called differential magnification, has been discussed by Blain (1999).
The magnification of a point source is expected to fall approximately
linearly with separation from a caustic line in the source plane.

Short of modelling the dust sources themselves, which is not
possible since images of the dust are not available, this problem can
be dealt with only approximately at best. Weak magnifications in
the optical mean that the quasar nucleus is relatively far from a caustic, in
which case the dust source will most likely get a comparable
overall boost to that of the optical.  At high optical magnifications, the boost for the
dust source will be lower.  Brian McLeod has provided us with model magnifications
for some of the quasars in this sample.  His results show that magnification
decreases with source size for high optical magnifications, and 
flattens out at about $m = 20$ for sources $\gtrsim 200$ pc. This is 
likely to be a minimum submillimeter/molecular source size (see Downes
et al.\ 1995; Downes et al.\ 1999; Kneib et al.\ 1997).   However, at 
magnifications $\lesssim 20$ the boost is essentially independent of 
source size.   This is illustrated in Figure 1.   

Based on this rough guidance, dust magnifications are set as follows:
for weak to moderate values, in the range $2.5-20$, the 
optical or radio core magnifications are used
(see Table 2); for magnifications greater than
20, 20 is adopted as the value; for sources with no available lensing
model, $m$ is set to 10.  This procedure, while inexact, will
generally provide 850\mic\ magnifications accurate to within a factor of
two or so for most sources, and should not be biased.

\subsection {Redshifts}

Of the 40 sources in the sample, 33 have measured redshifts.
All of these objects have broad emission lines and are therefore
classified as quasars.  Of the remaining 7 without detected 
emission lines, one is radio quiet and
likely to be a quasar, and 6 are radio loud and could be either quasars
or radio galaxies.  The mean of the measured redshifts is 2.4,
so for purposes of statistical analysis of the luminosities 
the unmeasured redshifts are set to this value, as indicated in Table 2.
Derived dust luminosities and masses (see below) are only weakly
dependent on $z$ because of the strong negative
K-correction that applies to submillimeter dust emission, so adopting
an intermediate redshift for the seven sources without redshifts 
should not greatly affect the statistical results.

\subsection {Corrected Fluxes, Luminosities, and Dust Masses}

Table 2 gives assumed magnifications and redshifts, corrected
850\mic\ fluxes (as described below), and calculated 850\mic\ dust
luminosities and dust masses.  Luminosities and masses are
provided for $H_0 = 50$ km s$^{-1}$ Mpc$^{-1}$, and two values
of the cosmological density parameter, $\Omega_0 = 1.0$ and 0.1.
These parameter values were chosen to facilitate direct comparison with the
recent 850\mic\  study of high redshift radio galaxies by Archibald
et al.\ (2001; hereinafter ADH01).  For simplicity it is assumed that
$\Omega_0 = \Omega_M$, the mass density parameter, and 
$\Omega_{\Lambda} = 0$; luminosity calculations with the  popular
$\Omega_M = 0.3$ and $\Omega_{\Lambda} = 0.7$ cosmology would lie
close to the $\Omega_0 = 0.1$ case for low redshifts, and 
roughly in the middle between $\Omega_0 = 0.1$ and 1.0 at high redshift.

For some of the radio loud objects the centimeter wavelength
synchrotron emission, extended to short wavelengths, can contaminate
the submillimeter fluxes.  In most cases sufficient data are available
at 3mm and 1.3mm to fit a synchrotron plus dust emission model.
A power law is assumed for the synchrotron, and an optically thin
isothermal dust emission template with $T_{dust} = 40$K and
emissivity index $\eta = 1.5$.  The fitted function is 
\begin{equation}
S_{\nu} = a\nu_r ^{\alpha} + b\nu_r ^{4.5}/[exp(h\nu_r/kT_{dust}) - 1], 
\end{equation}
where $\nu_r = (1+z)\nu$ is the frequency in the quasar rest frame.
This allows separation of the dust and synchrotron components and
calculation of the uncontaminated 850\mic\  flux.  The fit indicates all
synchrotron and no dust for B0712+472 and B1600+434, and 850\mic\ dust
upper limits are given for those sources in Table 2.  Significant dust
fluxes above the synchrotron power law were derived from the fits to
MG J0414+0534, B1608+656, and B1938+666.  The corrected fluxes for these
three objects are given in Table 2.  We estimate that that the correction
adds an additional uncertainty of about 20\% to the 850\mic\ fluxes of
MG J0414+0534 and B1938+666, and about 30\% for B1608+656.  These are added
in quadrature to the statistical give the uncertainties quoted in Table 2.

A small flux correction is required for F10214+4724, which was measured
at 800\mic\ (Rowan-Robinson et al.\ 1993) rather than 850\mic .  
The flux corrected to 850\mic , and listed in Table 2, is 42.7 mJy.
The dust template adopted above, with dust temperature $T_{dust} = 40$K 
and emissivity index $\eta = 1.5$, was used for the correction.

K-corrections for computing {\it rest frame} 850\mic\ luminosities are
also derived using this same dust template.
This is the template 
used by ADH01 (which see for rationale, discussion of
potential errors introduced by assuming the wrong template parameters, and
references).  To facilitate comparison with the radio galaxy results their
units of luminosity are also adopted, $L_{850}$ (W Hz$^{-1}$ sr$^{-1}$),
as is their method of  computing dust masses using \begin{equation}
M_{dust} = {S_{obs} D_L^2\over (1+z)\kappa_{\nu_r} B_{\nu_r}(T_{dust})},
\end{equation} where $S_{obs}$ is the observed flux density, $D_L$ is the
luminosity distance, $\kappa_{\nu_r}$ is the mass absorption coefficient
of the dust at $\nu_r$ (assumed to be equal to 0.15 m$^2$kg$^{-1}$
at 800\mic , and scaled to other wavelengths using $\lambda^{-1.5}$),
and $B_{\nu_r}(T_{dust})$ is the intensity of a blackbody at $\nu_r$
assuming isothermal emission from dust grains at temperature $T_{dust}$.

\section{Discussion}

Of the 40 sources presented in Table 1, 23 have clear detections of dust
emission (see Table 2). This detection rate of 58\% is almost twice that of
the large survey of high redshift radio galaxies of ADH01, which detected
14/46 radio galaxies (30\%) while going deeper in flux RMS (typically to 1
mJy) than the present survey and also using $2.0\sigma$ as the detection
level ($2.5\sigma$ is used here).  The difference in detection rate is
probably due to the lensing boost, rather than significant differences
between the dust luminosities of quasars and radio galaxies, as shown
below.  Another recent AGN survey, of $z \approx 4$ quasars by Omont et
al.\ (2001), had a 29\% detection rate in observations at 1.2mm that had
an average depth approximately equivalent to
ours at 850\mic\ (assuming a dust emissivity exponent $\eta$ of 1.5).

\subsection{Mean Luminosities}

Establishing the mean 850\mic\ dust luminosity and dust mass for the
lensed quasar sample, and investigating whether these quantities are
a function of redshift, is complicated by the presence of upper limits
(17 out of 40 measurements).  Extracting statistical information from
the data requires survival analysis techniques, and here we use the ASURV
software package developed by T.\ Isobe, M.\ LaValley, and E.\ Feigelson.
This package provides correlation analysis, estimates of mean values,
and linear regression parameters for samples containing both detections
and upper limits.

However, some caveats apply.  Estimation of the mean for flux-limited
samples using survival analysis is not very robust when the
upper limits are found mostly at the low end of the distribution,
tending to bias the estimation upward relative to the true value
as the percentage of upper limits increases.  Conversion of fluxes
to luminosities usually scrambles the upper limits because sources
are at varying distances, mitigating the bias effect when applied
to luminosities.  However, in this case the luminosities are nearly
independent of distance because of the steep negative K correction,
so the upper limits in luminosity, like those in flux, are also
largely clustered near the bottom.  

Here we analyze the lensed quasars in this sample, and compare
with the radio galaxies of ADH01.  For these two samples,
the radio galaxies would be more biased than the lensed quasars
because of the higher fraction of upper limits.  On the positive
side, the upper limits and detections for both samples are mixed to
some extent because of non-uniform flux limits and, for the lensed
quasars, the application of different boost factors.  We will proceed
with survival analysis with the recognition that the results probably
represent upper limits to the true mean luminosities.

That noted, the mean of a sample with upper limits can be computed using
the Kaplan-Meier estimator provided in ASURV.  Applying this to the quasar
sample, for $\Omega_0 = (1.0, 0.1)$, gives $<{\rm log}(L_{850})> = (22.47,
22.87)$ WHz$^{-1}$sr$^{-1}$, for 23 detections and 17 upper limits.
Using the data of ADH01 to calculate the mean luminosty for radio
galaxies, $<{\rm log}(L_{850})> = (22.82, 23.19)$ WHz$^{-1}$sr$^{-1}$,
for 14 detections and 32 upper limits.  At face value it would appear
that radio galaxies are more luminous than quasars in the submillimeter
by a factor of 2 or so, but the bias caused by the larger number of
upper limits in the radio galaxy sample could easily account for such a
difference.  The bias would be particularly severe if the luminosity
function is steep and has many more weak sources than strong ones, as
is almost certainly the case (see the 850\mic\ luminosity function
of local IRAS galaxies in Dunne et al.\ 2000).  A precise comparison
between the samples is further complicated by uncertainties in the lensing
magnifications of the quasars.  We conclude that the dust luminosities
of high-redshift quasars and radio galaxies are qualitatively similar.

\begin{figure*}
\rotate
\figurenum{2}
\epsscale{2.2}
\plottwo{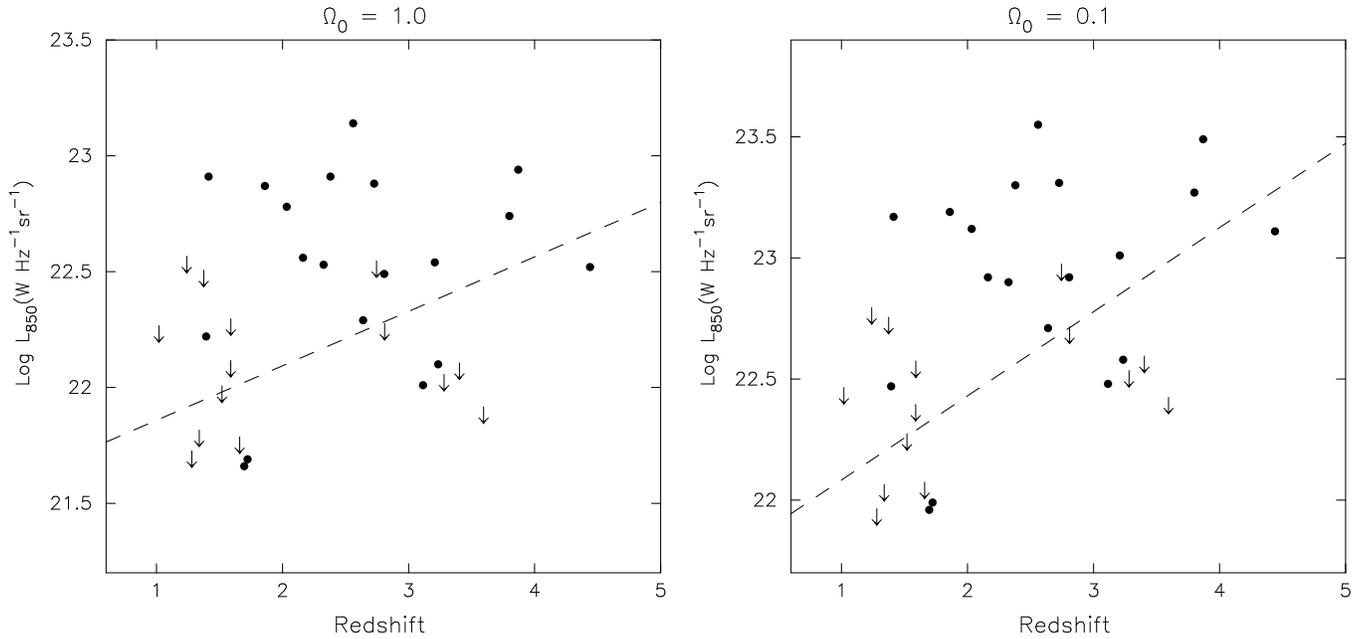}{fig2b.eps}
\caption{Monochromatic luminosity at 850\mic\ versus redshift for the two 
cosmologies considered.  The dashed lines represent best fit linear
regressions, taking upper limits into account using the Buckley-James
algorithm supplied in ASURV.  A plot with $\Omega_M = 0.3$ and $\Omega_{\Lambda}
= 0.7$ would appear intermediate between these two.  
Only sources with measured redshifts are plotted.
} \end{figure*}

How do the high redshift quasars compare with local objects?
From Table 2, the {\it detected} quasars have dust luminosities in
the range $21.66 < {\rm log}(L_{850}) < 23.14$ WHz$^{-1}$sr$^{-1}$
for $\Omega_0 = 1.0$ (for $\Omega_0 = 0.1$, the numbers are about
twice as high).  In the local universe, a number of
quasars  and Seyfert 1 galaxies have been observed at 850\mic , and
the luminosities of the detected ones typically fall within the
range of the detected members of the high-$z$ sample.  For example,
the luminosities of I Zw 1 (0050+124; $z = 0.061$), Mrk 1014 
(0157+001; $z = 0.163$),
and Mrk 376 (0710+457; $z = 0.056$) are: ${\rm log}(L_{850}) = 22.18$, 22.94,
and 22.06 respectively (using fluxes from Hughes et al.\ 1993).

A comparison can also be made with local (non-radio) galaxies.
Results are available from a SCUBA survey of a large sample of nearby
galaxies ($z < 0.07$) selected from the IRAS Bright Galaxy Sample.
Measurements of 104 galaxies with $S_{60} > 5.24$ Jy are reported by
Dunne et al.\ (2000); all were detected at 850\mic .  Roughly 65\%
are  bright enough (assuming $H_0 = 50$ km s$^{-1}$ Mpc$^{-1}$) 
to fall within
the definition of Luminous Infrared Galaxies (LIRGs), i.e., galaxies
with $L_{\rm IR} = L(8-1000\mu {\rm m}) > 10^{11} M_{\sun}$.\footnote{Dunne
et al.\ computed the far-infrared $40-1000$\mic\ values, $L_{FIR}$,
but given the steep mid-infrared galaxy spectra this approximates
the $8-1000$\mic\ luminosities within a few percent.}  A few would
be classified as ULIRGs with $L_{\rm IR} > 10^{12} M_{\sun}$, and the
rest have total infrared luminosities below $10^{11} M_{\sun}$.

In terms of monochromatic 850\mic\ luminosities, the range for the nearby
IRAS galaxies is $20.35 < {\rm log}(L_{850}) < 22.95$ WHz$^{-1}$sr$^{-1}$,
and the mean value is 22.15.\footnote{These values are independent of
$\Omega_0$ because the galaxies are nearby.} On average therefore
the quasars, with ${\rm log}(L_{850}) \sim 22.7$ WHz$^{-1}$sr$^{-1}$
(taking a value intermediate between $\Omega_0 = 0.1$ and 1), are
not too different from local infrared-bright galaxies: their mean
monchromatic 850\mic\ luminosity is a few times higher.

To estimate the total infrared luminosities of the quasars, 
a template spectrum is needed to scale from $L_{850}$ to $L_{IR}(8-1000\mu
{\rm m} )$.  The local quasar I Zw 1 has a well-measured
infrared/submillimeter spectrum (see Hughes et al.\ 1993) and
fits the bill -- its spectrum is flat in $\nu S_{\nu}$ throughout
the infrared and falls off sharply longward of 100\mic , typical
of most quasars.
This is in contrast to IRAS galaxies, which,
because of a sharp dropoff from 60\mic\ to 25\mic , have relatively
little power in the mid-infrared.   A I Zw 1--like
spectrum was integrated to derive an approximate conversion factor 
from $L_{850}$
to $L_{IR}(8-1000\mu {\rm m})$ of $1.2\times 10^{-10}$ $L_{\sun}$
(WHz$^{-1}$sr$^{-1})^{-1}$.  Using this, the estimated mean 8 to
1000\mic\ infrared luminosity of the quasars is $<L_{IR}$(8-1000\mic
)$>$ $\sim 6\times 10^{12}h_{50}^{-2} L_{\sun}$, where $h_{50}$
is the Hubble constant in units of 50 km s$^{-1}$ Mpc$^{-1}$.
This means that the bolometric infrared luminosities of quasars are
typically well into the range defined by ULIRGs.  The mean total
infrared luminosity of the local IRAS galaxies of Dunne et al.
(2000) is $1.3\times 10^{11}h_{50}^{-2} L_{\sun}$, about 45 times below 
that of the quasars.   Thus, while the submillimeter luminosities of
quasars are only a few times larger than IRAS galaxies, their total
infrared luminosities are much larger.  This is because quasars have
much flatter mid-infrared spectra, and substantially more power there.

\begin{figure*} 
\figurenum{3}
\includegraphics[height=15cm, angle=-90]{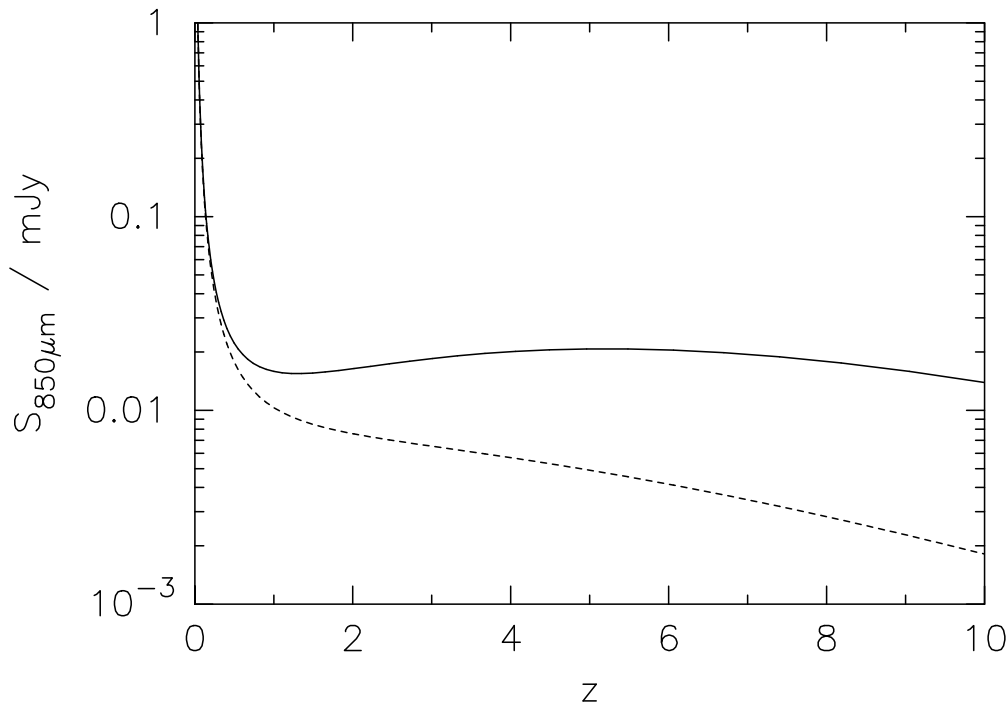} 
\caption{The observed 850\mic\ flux density as a function of redshift
for a submillimeter source having the dust emission template adopted in this
paper,
with $\eta = 1.5$ and $T_{dust} = 40$K.  The solid line is for an
$\Omega = 1.0$ universe, and the dashed line is for $\Omega = 0.1$.
Figure courtesy of E.N. Archibald.  } \end{figure*} \subsection{Relation
Between Luminosity and Redshift}

Figure 2 plots the quasar 850\mic\ luminosities versus redshift
for $\Omega_0 = 1.0$ and 0.1, including only sources with measured
redshifts. Ignoring the dashed regression lines for the moment, the
eye picks up a slight trend of increasing luminosity with redshift
for $\Omega_0 = 1$, and a definite trend for $\Omega_0 = 0.1$.
Because of the strongly negative K correction, observed fluxes for a
given submillimeter luminosity are essentially constant with redshift
above $z=1$ for $\Omega_0 = 1$, and slightly declining for $\Omega_0
= 0.1$ (see Figure 3).  

Thus the usual situation where flux-limited samples tend to pick up higher 
luminosity objects at high redshift does not apply to submillimeter-selected 
samples, or only weakly. The samples presented 
here, and by ADH01, were not 
selected at submillimeter wavelengths however. ADH01 mitigated against 
potential selection
effects at other wavelengths by selecting radio galaxies covering only 
a narrow range of radio 
luminosity. This approach was not possible for our sample of lensed 
quasars, though would naturally be adopted in any well-planned survey of 
field quasars. We must therefore consider the possibility that the observed 
trends with redshift are caused by sample anomalies. To this end, we have 
investigated possible correlations between optical/radio luminosity and 
redshift. No correlations are found, indicating that selection effects 
at other wavelengths are 
not responsible for the submillimeter trends.

The plotted regression lines take into account the
upper limits, and are derived using the Buckley-James method in ASURV.
They provide a stronger and more quantitative indication of a trend with
redshift than the eye alone, since at first glance the upper limits tend
to be interpreted as detections.  But since more of the upper limits
lie at low redshifts, they pull that end of the regression line down.

ASURV provides three test statistics for determining
the probability that a correlation exists within a given data set:
Cox's Proportional Hazard Model, Generalized Kendall's Tau, and
Generalized Spearman's Rho.  ASURV actually computes the probablility that
no correlation is present (the null hypothesis); the results are
listed in Table 3.  For $\Omega_0 = 1$, the probability of {\it no
correlation present} lies in the range $4-12\%$, depending on the
test.  The meaning of this for the Cox test, for example, is that 
in  96\% of randomly distributed samples the degree of correlation 
would be lower.  For $\Omega_0 = 0.1$ the correlation is stronger, 
and the null probability lies in the range $0.8 - 2.5\%$.  

Altering the dust template can change the degree of correlation and
the slope of the luminosity--redshift relation.  Lower temperatures and
smaller values of $\eta$ strengthen the relation, while higher temperatures
and larger values of $\eta$ weaken it.  For the worst case we consider
reasonable, with $\eta = 2$ and $T_{dust} = 60$K, the correlation
breaks down for $\Omega = 1$, but remains reasonably significant for
$\Omega = 0.1$.  The correlation probablilities for this template are also 
shown in Table 3.  However, evidence from other studies suggests
that $\eta$ is well under 2, and $T_{dust}$ is in the range $35-50$K
(Benford et al.\ 1999; Dunne et al.\ 2000; Ivison et al.\ 2000).

The total uncertainties in the submillimeter luminosity estimates for this
sample derive from the chosen cosmology, the applied magnification factor,
the synchrotron correction (for three sources), the K-correction (dust
template), and the statistical uncertainties in the flux measurements.
Uncertainties in the synchrotron correction have been accounted for in
the error estimates given in Table 2.  The chosen Hubble Constant gives
a constant offset, while the cosmological density parameter produces a
redshift dependence.  Computations with $\Omega = 0.1$ and 1.0 provide
the extremes of this dependence.  The K-correction also yields a redshift
dependence, which has been addressed by taking a very unfavorable (and
unlikely, we think) dust template and assessing the luminosity--redshift
relation for that case. 

Estimated uncertainties in luminosity for a
given source (relative to others in the sample) 
are probably roughly a factor of 3 when all is taken
into account. Some of the uncertainty is random (e.g., magnification) and some
is systematic.  We believe we have covered the reasonable
ranges of systematic error and conclude that the general trend of
submillimeter luminosity increasing with redshift is likely to be real.
However, given the uncertainties, the trend cannot be said to
be proven by the data.

Considering the redshift dependence of radio galaxies, the null 
probabilities for no correlation between $L_{850}$  and $z$
listed by ADH01 are $0.0-0.2\%$ for $\Omega_0 =
1$, and 0.00\% (all three statistics) for $\Omega_0 = 0.1$;  clearly
the statistical evidence, from survival analysis, of increasing
luminosity with redshift is stronger for the radio galaxies. This
does not mean that the {\it rate} of increase is stronger, but
rather that the scatter in the distribution is somewhat lower
for the radio galaxies than for the quasars in these samples.
For the $\Omega_0 = 0.1$ case, both distributions show roughly
an order of magnitude increase from $z = 1$ to $z = 4$. Most of
the increased probability of correlation for the radio galaxies
appears to be based on three quite luminous objects at $z \sim 4$
(see Figure 4 of ADH01).  We hypothesize that the larger scatter
in the quasar distribution is caused by two effects: the larger
number of detections, providing a broader probe of the luminosity
function, and uncertainties in the lensing corrections.

\subsection{Comparing Radio Loud and Radio Quiet Quasars}

Do radio loud and radio quiet quasars have different submillimeter
properties?  The Kaplan-Meier 
estimator has been applied to the radio loud and quiet luminosities 
separately, resulting in (for $\Omega = 1$)
$<L_{850}({\rm RLQ})> = 2.3\pm 0.7\times 10^{22}$ WHz$^{-1}$sr$^{-1}$
and $<L_{850}({\rm RQQ})> = 3.4\pm 0.7\times 10^{22}$ WHz$^{-1}$sr$^{-1}$,
with 9 detections and 8 upper limits for the RLQs and 14 detections
and 9 upper limits for the RQQs.  These values are consistent with each
other to within the errors, and we conclude that there is no evidence 
here for a systematic difference 
in submillimeter luminosity between radio loud and quiet quasars.  
This similarity in dust properties between optical and radio selected
quasars suggests that there is no dust-related difference between radio loud
and radio quiet AGN, as might be expected if one population resided in 
more massive systems than the other.  

Comparison between the quasar types for correlations with redshift was
not possible because of the small number of detected radio loud 
quasars having measured redshifts (5 detections out of 11 objects).

\section{Summary and Conclusions}

The present submillimeter survey of lensed quasars shows
that the dust
luminosities/masses of quasars are comparable to those of radio
galaxies (Archibald et al.\ 2001) in the same redshift range ($1 \lesssim
z \lesssim 4.4$), and that quasars, like radio galaxies, appear to have
increasingly powerful dust sources toward higher redshifts.  The mean
monochromatic 850\mic\ luminosity of high-redshift quasars is only a few
times higher than that of a local sample of IRAS-selected galaxies,
but because quasars have more power in the mid-infrared their total
infrared luminosities ($8-1000$\mic ) are substantially greater.  

The concordance of mean dust luminosity and luminosity evolution with
redshift between quasars and radio galaxies is broadly consistent
with, and supportive of, obscuration/orientation unified schemes
wherein these classes are by and large similar except for chance viewing
angle.  The submillimeter dust flux is likely to be optically
thin and therefore orientation independent.  The presence or
absence of a strong radio source appears to have little effect
on the submillimeter power once synchroton contamination has been
accounted for, and this argues against large differences in the total
masses of radio quiet versus radio loud systems.

The tentative evidence presented here for higher dust luminosities and 
inferred dust masses at high redshift could have 
several interpretations.  One possibility is that the increase is illusory, 
for example if dust properties change over cosmic time in 
such a way as to cause 
an increased emissivity per unit dust mass at high $z$, or if 
the spectral energy distribution of grain emission changes with time 
causing the {\it derived} relative luminosities/masses to change.  
The template SED used here is not necessarily the 
correct one in general, and cannot be correct for all specific objects  
since galaxy studies have shown  
a fair amount of variation in the far-IR/submillimeter 
SEDs from one object to the next (see, e.g., Figure 2 of Adelberger \& Steidel
2000).  Other SED shapes can dampen the trend with redshift (while
yet others can strengthen it).
Greenberg \& Shen (2000) have stressed that 
dust in the early universe could be much different than it is 
today, but the form of any cosmic evolution of mean dust 
properties from $z=4$ to $z=1$ is anybody's guess.

Dust sources in quasars are powered by  
absorption of energy from the central AGN and by star formation, 
with the proportionate mix between the two being in general unknown.  
An increased
mass of dust in early galaxies could come about because of more frequent
galaxy interactions and mergers when the universe was smaller and more dense. 
These mergers could, in addition to dumping mass into the central regions of
the host galaxy, trigger vigorous star formation to power the dust.

Alternatively, if the primary energy source for the grains is the
central AGN, the increased luminosity 
could be caused by a larger dust covering factor at higher $z$, meaning
that more of the quasar light would be intercepted and reprocessed into  
IR/submillimeter emission.  A larger covering factor could  
result from a larger total dust mass, or simply a larger fraction of the 
dust being exposed
to the primary radiation.  If there is a torus surrounding the AGN, it 
could be physically thicker (i.e., smaller opening angle).  A second
possibility is that material has not had a chance to settle into an 
optically thick torus 
because of the youth of the systems and their more violent 
environments, leaving relatively more of the grains exposed to 
the nuclear radiation.  The larger covering factor interpretation imples 
a larger fraction of obscured AGNs at high redshifts, 
as a lower proportion of direct quasar light escapes to distant observers.  

Models for the x-ray background are still uncertain, but 
recent work by Gilli et al.\ (2001) finds a better fit to the currently 
available x-ray 
data with an increasing ratio of obscured to unobscured AGNs toward higher 
redshifts.  That particular model 
seems to indicate that the evolution stops above about $z = 1.5$.
The increasing dust luminosity with redshift found 
by Archibald et al.\ (2001) for radio galaxies, and the similar relation
found here for quasars, suggests that the 
proportion of obscured AGNs may continue to increase out to $z = 2$ or
beyond.




\acknowledgments

The authors thank the JCMT for generous allocations of telescope
time for this project, and Iain Coulson for help with some of the 
observations.  Frank Bertoldi kindly obtained some of the IRAM 1.3mm 
bolometer data.  We thank 
Joseph Lehar and Brian McLeod for providing 
lensing magnifications from the Castles project,
and Ski Antonucci and Andrew Blain for comments on the manuscript.
Brian McLeod contributed substantially to our understanding of magnification
effects by running magnification 
versus source size models for use in Figure 1.
Eric Fiegelson kindly provided advice on the use of ASURV and interpretation 
of its output, and Elese Archibald granted permission to use Figure 3.
We are grateful to the referee for helpful suggestions.
The CASTLeS web page, http://cfa-www.harvard.edu/castles, was used 
extensively, as was the NASA Extragalactic Database (NED).

\newpage
\begin{deluxetable}{lclccccc}
\tablenum{1}
\tabletypesize{\scriptsize}
\setlength{\tabcolsep}{0.19in} 
\tablewidth{0pt}
\footnotesize
\tablecaption{mm/Sub-mm Measurements of Lensed Quasars}
\tablehead{
 \colhead{Source}
&\colhead{Type}
&\colhead{$z$}
&\colhead{$S_{3000}$}
&\colhead{$S_{1300}$}
&\colhead{$S_{850 }$}
&\colhead{$S_{450 }$}
&\colhead{Notes}
\\
 \colhead{}
&\colhead{}
&\colhead{}
&\colhead{(mJy)}
&\colhead{(mJy)}
&\colhead{(mJy)}
&\colhead{(mJy)}
&\colhead{}
}
\startdata
0047$-$2808       &RQQ  &3.595& $< 2.0$   &  ...      &$< 7.0$       & ...       &1\\
CLASS B0128+437   &RLQ  &-----&   ...     &  ...      &$< 6.0$       & ...       &\\
UM673             &RQQ  &2.727& $< 1.3$   &$< 5.8$    & 12.0 (2.2)   & $< 40$    &\\
HE 0230$-$2130    &RQQ  &2.162&   ...     &  ...      &  21.0 (1.7)  & 77 (13)   &\\
MG J0414+0534     &RLQ  &2.639& 40 (2.0)  & 20.7 (1.3)&  25.3 (1.8)  & 66 (16)   &2\\
CLASS B0712+472   &RLQ  &1.339& 22 (0.4)  & 9.8  (1.4)&   7.4 (1.8)  & ...       &8,10\\
CLASS B0739+367   &RLQ  &-----&   ...     & $<2.8$    &   6.6 (1.3)  & 36 (10)   &2\\
MG J0751+2716     &RLQ  &3.208& 4.1 (0.5) & 6.7 (1.3) &25.8 (1.3)    & 71 (15)   &1\\
HS 0818+1227      &RQQ  &3.115&   ...     &  ...      & 4.6 (1.7)    & $<83$     &\\
APM 08279+5255    &RQQ  &3.870&   ...     & 24 (2)    &  84  (3)     &285 (11)   &3\\
SBS 0909+532      &RQQ  &1.375& $< 0.8$   & $< 2.3$   &$< 5.5$       & ...       &2\\
RX J0911.4+0551   &RQQ  &2.807& 1.7 (0.3) &10.2 (1.8) &  26.7 (1.4)  & 65 (19)   &1\\
RX J0921+4529     &RQQ  &1.66 &   ...     &  ...      &$< 4.3$       & ...       &\\
FBQS J0951+2635   &RQQ  &1.24 &   ...     & $< 2.5$   &$< 3.8$       & ...       &2\\
BRI 0952$-$0115   &RQQ  &4.440&   ...     & $< 2.3$   &13.4 (2.3)    & ...       &2,4\\
0957+561          &RLQ  &1.413&   14.2    & $< 4.0$   & 7.5 (1.4)    & ...       &5\\
LBQS 1009$-$0252  &RQQ  &2.746&   ...     & $< 5.0$   & $<4.5$       & $<58$     &2\\
IRAS F10214+4724  &RQQ  &2.380&   ...     & 24   (5)  &  50  (5)     &273 (45)   &6\\
HE 1104$-$1805    &RQQ  &2.326& $< 5.0$   &  5.3 (0.9)&  14.8 (3.0)  & ...       &2\\
PG 1115+080       &RQQ  &1.723& $< 0.8$   &$< 3.0$    & 3.7 (1.3)    & ...       &1\\
CLASS B1127+385   &RLQ  &-----&   ...     &$< 2.8$    & 13.9 (2.3)   & $<65$     &2\\
CLASS B1152+199   &RLQ  &1.019&   ...     &  ...      & $ < 6.5$     & $<70$     &\\
1208+1011         &RQQ  &3.8  & $< 0.9$   &  2.8 (0.9)& 8.1 (2.0)    & ...       &2\\
HST 12531$-$2914  &RQQ  &-----&   ...     &  ...      & $< 6.4$      & ...       &\\
CLASS B1359+154   &RLQ  &3.235&   ...     &  ...      & 11.5 (1.9)   & 39 (10)   &\\
HST 14113+5211    &RQQ  &2.81 &   ...     &  ...      & $< 3.6$      & ...       &\\
H 1413+117        &RQQ  &2.560&   ...     & 18.0 (2.0)&   58.8(8.1)  &224 (38)   &2,7\\
HST 14176+5226    &RQQ  &3.403& $< 0.9$   & $< 4.0$   & $< 3.5$      & $< 26$    &2\\
SBS 1520+530      &RQQ  &1.860& $< 0.6$   &  2.8 (0.8)& 9.4 (2.6)    & ...       &2\\
CLASS B1555+375   &RLQ  &1.59 &   ...     &  ...      & $< 6.8$      & $< 39$    &\\
CLASS B1600+434   &RLQ  &1.589& 25 (0.3)  &12.6 (2.3) &  7.3 (1.8)   &           &1,10\\
CLASS B1608+656   &RLQ  &1.394& 8.1 (0.4) &  5.6 (1.7)&   8.1 (1.7)  &           &9\\
FBQS J1633+3134   &RQQ  &1.520&   ...     &  ...      & $< 3.5$      & ...       &\\
CLASS B1938+666   &RLQ  &-----&   ...     & 14.7 (2.0)&  34.6 (2.0)  &126 (22)   &9\\
MG J2016+112      &RLQ  &3.282& 1.8 (0.2) &$< 2.5$    & $< 4.8$      & ...       &1\\
CLASS B2045+265   &RLQ  &1.280&   ...     &  ...      & $< 3.7$      & $< 22$    &\\
CLASS B2114+022   &RLQ  &-----&   ...     &  ...      & $< 4.3$      & ...       &\\
HE 2149$-$2745    &RQQ  &2.033& $< 6.8$   &  ...      & 8.0 (1.9)    & ...       &1\\
2237+0305         &RQQ  &1.696& $< 0.8$   &  ...      & 2.8 (0.9)    & $< 17$    &1\\
CLASS B2319+051   &RLQ  &-----&   ...     &  $<3$     & 3.9 (1.2)    & 40 (8)    &2\\
\enddata
\tabnotes{
(1)  IRAM PdBI observations at both 3mm and 1mm (where reported).
(2)  IRAM PdBI at 3mm (where reported), IRAM 30m telescope at 1.3mm.
(3)  Also detected by Lewis et al.\ (1998) at 850 and 450\micron , our measurements reported here; 
1.3mm flux from Lewis et al.\ (1998).
(4)  Also detected at 850\mic\ by McMahon et al.\ (1999).  
(5)  3mm flux from Planesas et al.\ (1999).  1.3mm flux from IRAM 30m.
(6)  All fluxes from Rowan-Robinson et al.\ (1993); measurements at 1100\mic , 800\mic , and
450\mic .
(7)  450\mic\ flux from Barvainis et al.\ (1992, 1995); 850\mic\ flux from this work.
(8)  1.3mm measurement is weighted average of fluxes from IRAM PdBI and 30m telescope.  3mm flux is from PdBI.
(9) 1.3mm flux from JCMT using SCUBA.  3mm flux from PdBI (where reported).
(10) Consistent with pure synchrotron.
}

\end{deluxetable}

\newpage
\begin{deluxetable}{lccccccccc}
\tablenum{2}
\tabletypesize{\scriptsize}
\setlength{\tabcolsep}{0.06in} 
\tablewidth{0pt}
\footnotesize
\tablecaption{850\mic\ Luminosities and Dust Masses}
\tablehead{
&&&&&\multicolumn{2}{c}{$\Omega_0=1.0, H_0=50$}&\multicolumn{2}{c}{$\Omega_0=0.1, H_0=50$}&\\
\colhead{Source}
&\colhead{~~~$z$\tablenotemark{a}~~~}
&\colhead{$S_{850}$\tablenotemark{b}}
&\colhead{$m$\tablenotemark{c}}
&\colhead{$m$\tablenotemark{d}}
&\colhead{log~$L_{850}$}
&\colhead{log~$M_{dust}$}
&\colhead{log~$L_{850}$}
&\colhead{log~$M_{dust}$}
&\colhead{Refs\tablenotemark{e}}
\\
\colhead{}
&\colhead{}
&\colhead{(mJy)}
&\colhead{(model)}
&\colhead{(adopted)}
&\colhead{(WHz$^{-1}$sr$^{-1}$)}
&\colhead{($M_{\sun}$)}
&\colhead{(WHz$^{-1}$sr$^{-1}$)}
&\colhead{($M_{\sun}$)}
&\colhead{}
}
\startdata
0047$-$2808       &  3.595    &  $< 7.0$  & 22.4& 20.0&  $<21.88$&  $< 7.35$&  $<22.39$&   $<7.86$&1\\
CLASS B0128+437       &{\it 2.400}&  $< 6.0$  &  ...& 10.0&  $<22.15$&  $< 7.63$&  $<22.55$&   $<8.02$& \\ 
UM673             &  2.727    & 12.0(2.2) &  3.7&  3.7&  ~~~22.88&   ~~~8.35&  ~~~23.31&   ~~~8.77&1\\
HE 0230$-$2130    &  2.162    & 21.0 (1.7)& 14.5& 14.5&  ~~~22.56&   ~~~8.03&  ~~~22.92&   ~~~8.39&1\\
MG J0414+0534     &  2.639    & 16.7 (3.8)& 26.9& 20.0&  ~~~22.29&   ~~~7.76&  ~~~22.72&   ~~~8.18&1\\
CLASS B0712+472       &  1.339    &  $< 4.5$  & 50.1& 20.0&  $<21.78$&   $<7.25$&  $<22.03$&   $<7.50$&1\\
CLASS B0739+667       &{\it 2.400}&  6.6 (1.3)&  ...& 10.0&  ~~~22.20&   ~~~7.67&  ~~~22.59&   ~~~8.06& \\ 
MG J0751+2716     &  3.208    & 25.8 (1.3)& 16.6& 16.6&  ~~~22.54&   ~~~8.01&  ~~~23.01&   ~~~8.48&1\\ 
HS 0818+1227      &  3.115    &  4.6 (1.7)&  ...& 10.0&  ~~~22.01&   ~~~7.48&  ~~~22.48&   ~~~7.95& \\ 
APM 08279+5255    &  3.870    & 84.0 (3.0)& 100 & 20.0&  ~~~22.94&   ~~~8.42&  ~~~23.49&   ~~~8.96&2\\ 
SBS 0909+532      &  1.375    &  $<5.5$   &  5.0&  5.0&  $<22.47$&   $<7.94$&  $<22.72$&   $<8.19$&1\\ 
RX J0911.4+0551   &  2.807    & 26.7 (1.4)& 21.8& 20.0&  ~~~22.49&   ~~~7.96&  ~~~22.92&   ~~~8.39&1\\ 
RX J0921+4528     &  1.660    &  $<4.3$   & 23.2& 20.0&  $<21.75$&   $<7.22$&  $<22.04$&   $<7.51$&1\\ 
FBQS J0951+2635    &  1.240    &  $<3.8$   &  3.0&  3.0&  $<22.53$&   $<8.00$&  $<22.76$&   $<8.22$&1\\ 
BRI 0952$-$0115   &  4.440    & 13.4 (2.3)&  8.3&  8.3&  ~~~22.52&   ~~~7.99&  ~~~23.11&   ~~~8.57&1\\ 
0957+561          &  1.413    &  7.5 (1.4)&  2.5&  2.5&  ~~~22.91&   ~~~8.39&  ~~~23.17&   ~~~8.64&1\\ 
LBQS 1009$-$0252   &  2.746    &  $<4.5$   &  3.2&  3.2&  $<22.51$&   $<7.98$&  $<22.94$&   $<8.40$&1\\ 
IRAS F10214+4724     &  2.380    & 42.7 (5.0)&100  & 13.0&  ~~~22.91&   ~~~8.38&  ~~~23.30&   ~~~8.76&3\\
HE 1104$-$1805    &  2.326    & 14.8 (3.0)& 10.8& 10.8&  ~~~22.53&   ~~~8.00&  ~~~22.90&   ~~~8.37&1\\
PG 1115+080       &  1.723    &  3.7 (1.3)& 27.7& 20.0&  ~~~21.69&   ~~~7.16&  ~~~21.99&   ~~~7.46&1\\ 
CLASS B1127+385       &{\it 2.400}& 13.9 (2.3)&  3.0&  3.0&  ~~~23.05&   ~~~8.52&  ~~~23.44&   ~~~8.91&1\\ 
CLASS B1152+199       &  1.019    &  $<6.5$   &  ...& 10.0&  $<22.23$&   $<7.71$&  $<22.43$&   $<7.90$& \\ 
1208+1011         &  3.800    &  8.1 (2.0)&  3.1&  3.1&  ~~~22.74&   ~~~8.21&  ~~~23.27&   ~~~8.74&1\\ 
HST 12531$-$2914  &{\it 2.400}&  $<6.4$   &  7.8&  7.8&  $<22.29$&   $<7.77$&  $<22.68$&   $<8.15$&1\\ 
CLASS B1359+154       &  3.235    & 11.5 (1.9)&  118& 20.0&  ~~~22.10&   ~~~7.57&  ~~~22.58&   ~~~8.05&1\\ 
HST 14113+5211    &  2.810    &  $<3.6$   &  4.8&  4.8&  $<22.24$&   $<7.71$&  $<22.68$&   $<8.14$&1\\ 
H 1413+117        &  2.560    &58.8 (8.1) &  9.6&  9.6&  ~~~23.14&   ~~~8.61&  ~~~23.55&   ~~~9.02&1\\ 
HST 14176+5226    &  3.403    &  $<3.5$   &  6.6&  6.6&  $<22.07$&   $<7.54$&  $<22.56$&   $<8.03$&1\\ 
SBS 1520+530      &  1.860    &  9.4 (2.6)&  3.3&  3.3&  ~~~22.87&   ~~~8.34&  ~~~23.19&   ~~~8.66&1\\ 
CLASS B1555+375        &  1.59     &  $<6.8$   &  ...& 10.0&  $<22.22$&   $<7.69$&  $<22.61$&   $<8.07$& \\ 
CLASS B1600+434       &  1.589    &  $<4.5$   &  ...& 10.0&  $<22.08$&   $<7.55$&  $<22.36$&   $<7.83$& \\ 
CLASS B1608+656       &  1.394    &  6.6 (2.6)& 10.8& 10.8&  ~~~22.22&   ~~~7.69&  ~~~22.47&   ~~~7.94&1\\ 
FBQS J1633+3134    &  1.520    &  $<3.5$   &  ...& 10.0&  $<21.97$&   $<7.44$&  $<22.24$&   $<7.71$& \\ 
CLASS B1938+666       &{\it 2.400}& 32.0 (6.7)&173  & 20.0&  ~~~22.59&   ~~~8.06&  ~~~22.98&   ~~~8.44&1\\ 
MG J2016+112      &  3.282    &  $<4.8$   &  ...& 10.0&  $<22.02$&   $<7.49$&  $<22.50$&   $<7.97$& \\ 
CLASS B2045+265       &  1.280    &  $<3.7$   & 81.2& 20.0&  $<21.69$&   $<7.16$&  $<21.93$&   $<7.40$&1\\ 
CLASS B2114+022       &{\it 2.400}&  $<4.3$   &  ...& 10.0&  $<22.01$&   $<7.48$&  $<22.40$&   $<7.87$& \\ 
HE 2149$-$2745    &  2.033    &  8.0 (1.9)&  3.4&  3.4&  ~~~22.78&   ~~~8.25&  ~~~23.12&   ~~~8.59&1\\ 
2237+0305         &  1.696    &  2.8 (0.9)& 16.0& 16.0&  ~~~21.66&   ~~~7.14&  ~~~21.96&   ~~~7.43&1\\ 
CLASS B2319+051       &{\it 2.400}&  3.9 (1.2)&  ...& 10.0&  ~~~21.98&   ~~~7.45&  ~~~22.36&   ~~~7.83& \\ 
\enddata
\tablenotetext{a} {Redshifts in italics are assumed, based on mean of the known redshifts for this sample.}
\tablenotetext{b} {850\mic\ fluxes corrected in 3 cases (MG J0414+0534, 
B1608+656, and B1938+666) for synchtrotron 
contamination, or a small difference in wavelength from 850\mic\ (F10214+4724). Quoted 
uncertainties have been increased to account for the uncertainties in the
synchrotron corrections; 
see Section 2.4 of the text.}
\tablenotetext{c} {Lensing magnification based on models of the optical or centimeter radio emission.  
See footnote (e) for references.}
\tablenotetext{d} {Adopted magnifications: if there is no model available, $m$ is set to 10.0; 
if the model gives $m>20$, $m = 20.0$ is assumed (with the exception of F10214+4724; see footnote (e)).  
See $\S 2.2$ for discussion.}
\tablenotetext{e} {References for the magnifications are (1) J. Lehar \& B. McLeod (2001, private
communication); (2) Egami et al.\ (2000);
(3) optical value from Eisenhardt et al. (1996), adopted value based on submillimeter estimate from 
Downes et al.\ (1995).}
\end{deluxetable}

\begin{deluxetable}{lcccc}
\tablenum{3}
\setlength{\tabcolsep}{0.16in}
\tablewidth{0pt}
\footnotesize
\tablecaption{$L_{850}$ vs. Redshift Correlation Results}
\tablehead{
&\multicolumn{4}{c} {Probability of No Correlation Present}\\
&\multicolumn{2}{c} {\it Nominal case}
&\multicolumn{2}{c} {\it Worst case}\\
&\multicolumn{2}{c} {$\beta=1.5$, $T=40$K}
&\multicolumn{2}{c} {$\beta=2.0$, $T=60$K}\\
\colhead{ASURV Test}
&\colhead{$\Omega_0 = 1.0$}
&\colhead{$\Omega_0 = 0.1$}
&\colhead{$\Omega_0 = 1.0$}
&\colhead{$\Omega_0 = 0.1$}
}
\startdata
Cox Proportional       & 4.1\%   &  0.8\% & 20\% & 4.1\%  \\
Kendall's Tau          & 11.9\%  &  2.5\% & 39\% & 10\%   \\
Spearman's Rho         & 6.9\%   &  1.3\% & 26\% & 6.5\%  \\
\enddata
\end{deluxetable}

\end{document}